\documentclass[aps,twocolumn,showpacs,preprintnumbers,prb]{revtex4}

\usepackage{graphicx}
\usepackage{dcolumn}
\usepackage{bm}

\begin{document}


\title{Comment on "Diffusion Monte Carlo study of jellium surfaces: Electronic
densities and pair correlation functions"}

\author{J. M. Pitarke }
\affiliation{Materia
Kondentsatuaren Fisika Saila, Zientzi Fakultatea, Euskal Herriko
Unibertsitatea,\\ 644 Posta kutxatila, E-48080 Bilbo, Basque Country\\
and Donostia International Physics Center (DIPC) and Centro
Mixto CSIC-UPV/EHU,\\
Manuel de Lardizabal Pasealekua,
E-20018 Donostia, Basque Country}

\date{11 August 2003}

\begin{abstract}
In a fixed-node diffusion Monte Carlo calculation of the total energy of
jellium slabs, Acioli and Ceperley [Phys. Rev. B {\bf 54}, 17199 (1996)]
reported jellium surface energies that at low electron densities were
significantly higher than those predicted in the local-density approximation
(LDA) of density-functional theory. Assuming that the fixed-node error
in the slab and the bulk calculations cancel out, we show that their data
yield surface energies that are considerably closer to the LDA and in
reasonable agreement with those obtained in the random-phase approximation.
\end{abstract}


\maketitle

Acioli and Ceperley\cite{acioli} presented the results of fixed-node difussion
Monte Carlo (DMC) calculations of the total energy of jellium slabs at five
different electron densities. Assuming that the released-node correction
is very small at the electron densities of interest, these authors extracted
surface energies by substracting from the fixed-node slab energy the
corresponding released-node bulk energies of Ceperley and Alder,\cite{ca} as
parametrized by Perdew and Zunger.\cite{zunger}  They concluded that at low
electron densities jellium surface energies are significantly higher than
those predicted in the local-density approximation (LDA) of density-functional
theory (DFT).

In this Comment, we show that combining fixed-node slab and release-node bulk
energies results in substantial imprecision. Instead, we expect a large degree
of cancellation between the fixed-node error in the slab and the bulk
calculations, and conclude that by substracting from the fixed-node slab
energies the corresponding fixed-node bulk energies one obtains jellium surface
energies that are considerably closer to the LDA\cite{zhang} and in reasonable
agreement with those obtained in the random-phase approximation
(RPA).\cite{pitarke1}

Acioli and Ceperley\cite{acioli} considered jellium slabs of density
$n_0=3/4\pi\,(r_s\,a_0)^3$ ($a_0$ is the Bohr radius), the thickness of the
positive background in each slab being $L=7.21\,(r_s\,a_0)$.
Surface energies were then obtained from
\begin{equation}\label{sigma}
\sigma=\left[\varepsilon^{slab}-\varepsilon^{bulk}\right]{n_0\,L\over 2},
\end{equation}
where $\varepsilon^{slab}$ and $\varepsilon^{bulk}$ represent slab and bulk
energies per particle, respectively. Fixed-node energies per particle in bulk
jellium were reported by Ceperley\cite{ceperley} to be higher than their
release-node counterparts by $9\times 10^{-4}\,{\rm Ry}$ for $r_s=2$ and
$2\times 10^{-4}\,{\rm Ry}$ for $r_s=5$. Hence, combining fixed-node slab and
release-node bulk energies Eq.~(\ref{sigma}) yields for $r_s=2$ and $5$ surface
energies that are
too large by $\sim 150$ and $5\,{\rm erg/cm^2}$, respectively. These
fixed-node errors represent $\sim 40\%$ and $\sim 20\%$ of the LDA
correlation energy for $r_s=2$ and $5$, respectively. Furthermore, they are
comparable to the deviation of the total surface energies reported in
Ref.~\onlinecite{acioli} from those obtained in the LDA.

In order to derive reliable surface energies from the fixed-node slab
calculations of Ref.~\onlinecite{acioli}, we require fixed-node energies of
bulk jellium. Such calculations were reported by Ceperley\cite{ceperley} and
more recently by Kwon {\it et al.}\cite{kwon} for $r_s$=1, 5, 10, and 20. The
fixed-node Slater-Jastrow correlation energies of Kwon {\it et al.}\cite{kwon}
can be parametrized in Perdew-Zunger\cite{zunger} form, as
follows\cite{perdew1}
\begin{equation}\label{par}
\varepsilon_c^{bulk}={-0.32172\over
1+1.3606\,\sqrt{r_s}+0.3391\,r_s}\,{\rm Ry}\quad (r_s\ge 1),
\end{equation}
or in Perdew-Wang\cite{pw} form, as follows
\begin{equation}\label{par2}
\varepsilon_c^{bulk}=-(0.12436+0.027404\,r_s)\,\ln\left(1+\frac{16.082}
{\alpha}\right)\,{\rm Ry},
\end{equation}
with
\begin{equation}
\alpha=7.5957\,r_s^{1/2}+3.5876\,r_s+1.8207\,r_s^{3/2}+0.47746\,r_s^2.
\end{equation}
Both Eqs.~(\ref{par}) and (\ref{par2}) have been adjusted to fit the fixed-node
correlation energies of Kwon {\it et al.}\cite{kwon} for $r_s$=1, 5, and 10.
The fixed-node bulk energy $\varepsilon^{bulk}=-0.01482\,{\rm Ry}$ obtained by
Ceperley and Alder\cite{ca2} for $r_s=2.07$ is underestimated by
Eq.~(\ref{par}) by $2\times 10^{-5}\,{\rm Ry}$ and overestimated by
Eq.~(\ref{par2}) by $2\times 10^{-4}\,{\rm Ry}$. At $r_s=20$, Eqs.~(\ref{par})
and (\ref{par2}) underestimate the fixed-node bulk energy of Kwon {\it et
al.}\cite{kwon} by $2\times 10^{-4}\,{\rm Ry}$ and $10^{-4}\,{\rm Ry}$,
respectively.

\begin{table}
\caption{Surface energies of jellium slabs with $L=7.21\,(r_s\,a_0)$, as
obtained from Eq.~(\ref{sigma})
by combining the Acioli-Ceperley fixed-node slab energies with the fixed-node
bulk energies of either Eq.~(\ref{par}) ($\sigma_1$) or Eq.~(\ref{par2})
($\sigma_2$), and by combining the Acioli-Ceperley fixed-node slab energies
with release-node bulk energies ($\sigma_{AC}$). At $r_s=2.07$, combining the
Acioli-Ceperley fixed-node slab energy with the corresponding fixed-node bulk
energy $\varepsilon^{bulk}=-0.01482\,{\rm Ry}$ of Ceperley and Alder\cite{ca2}
yields $\sigma=-558$. $\sigma_{\rm LDA}$ and $\sigma_{\rm RPA}$ represent LDA
and RPA surface energies. $\sigma_0=\sigma_s+\sigma_{es}+\sigma_x$ represents
the combined kinetic ($\sigma_s$), electrostatic ($\sigma_{es}$), and exchange
($\sigma_x$) contributions to the total surface energy $\sigma$. The
correlation surface energy is simply $\sigma_c=\sigma-\sigma_0$. Units are
erg/cm$^2$ ($1\,{\rm erg}/{\rm cm}^2=6.2415\times 10^{-5}\,{\rm
eV}/{\rm\AA}^2)$.}
\begin{ruledtabular} \begin{tabular}{lcccccc}
$r_s$&$\sigma_0$&$\sigma_1$&$\sigma_2$&$\sigma_{\rm AC}$ &$\sigma_{\rm
LDA}$&$\sigma_{RPA}$\\ \hline
1.87&-2402&-1197&-1247&-1035&-1557&-1424\\
2.07&-1234&-554&-592&-416&-600&-497\\
2.66&-131&242&226&327&180&233\\
3.25&52&312&305&365&227&258\\
3.93&72&251&249&284&173&191\\
\end{tabular} \end{ruledtabular} \label{table1}
\end{table}

Adding to the correlation energy of Eqs.~(\ref{par}) and (\ref{par2}) the
well-known kinetic and exchange energies of a uniform electron gas\cite{pines}
and combining these energies with the fixed-node slab energies of Acioli and
Ceperley,\cite{acioli} Eq.~(\ref{sigma}) yields the surface energies
($\sigma_1$ and $\sigma_2$) shown in Table~I. Also shown in this Table are the
surface energies
reported in Ref.~\onlinecite{acioli} ($\sigma_{\rm AC}$), together with the
LDA and RPA surface energies that we have obtained for jellium slabs with
$L=7.21\,(r_s\,a_0)$. Small differences of no more than $10\,{\rm erg/cm^2}$
between
these LDA and RPA surface energies and those reported before for the
semi-infinite jellium\cite{zhang,pitarke1} are due to the finite size of our
jellium slabs.

An inspection of Table~I shows that using in Eq.~(\ref{sigma}) the fixed-node
bulk energies of either Eq.~(\ref{par}) or Eq.~(\ref{par2}) brings the DMC
surface energies closer to the LDA and to reasonable agreement with the RPA.
This is consistent with recent
work, where it was shown that upon surface formation there is a
persistent cancellation of short-range correlation effects beyond the
RPA.\cite{pitarke2} Other approaches have also led to the conclusion that
the actual jellium surface energies should be only slightly higher than those
obtained in the LDA.\cite{x,y,z}

Li {\it et al.}\cite{li} calculated the fixed-node DMC surface energy of a
jellium slab with $r_s=2.07$, the thickness of the positive background being
$L=8.52\,(r_s\,a_0)$. These authors,\cite{li} unlike Acioli and
Ceperley,\cite{acioli}
extracted from their fixed-node slab energy the corresponding fixed-node bulk
energy ($\varepsilon^{bulk}=-0.01482\,{\rm Ry}$),\cite{ca2} and found
$\sigma=-465\,{\rm erg/cm^2}$;~\cite{note0} they also performed LDA
calculations and found an LDA surface energy of $-567\,{\rm erg/cm^2}$. We
have calculated the LDA and RPA surface energies of a jellium slab with
$r_s=2.07$ and $L=8.52\,(r_s\,a_0)$, and have found $\sigma_{\rm
RPA}=-485\,{\rm
erg/cm^2}$ and $\sigma_{LDA}=-589\,{\rm erg/cm^2}$,~\cite{note1} which are both
only $\sim 20\,{\rm erg/cm^2}$ smaller than the corresponding DMC surface
energy reported in
Ref.~\onlinecite{li}.

\begin{figure}
\includegraphics[width=0.45\textwidth,height=0.3375\textwidth]{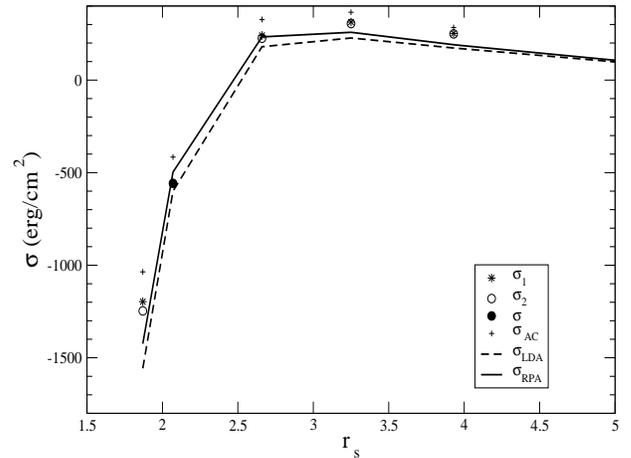}
\caption{Surface energies of jellium slabs with $L=7.21\,(r_s\,a_0)$, as
obtained
from Eq.~(\ref{sigma}). Stars and open circles represent DMC surface energies
obtained by
combining the Acioli-Ceperley fixed-node slab energies with the fixed-node bulk
energies of Eqs.~(\ref{par}) and (\ref{par2}), respectively. The solid circle
represents the DMC surface energy obtained at $r_s=2.07$ by combining the
Acioli-Ceperley fixed-node slab energy with the corresponding fixed-node bulk
energy $\varepsilon^{bulk}=-0.01482\,{\rm Ry}$ of Ceperley and
Alder.\cite{ca2}
Crosses represent the surface energies obtained in
Ref.~\onlinecite{acioli} by combining fixed-node slab and
release-node bulk energies. Dashed and
solid lines represent LDA and RPA calculations, respectively.} \label{fig1}
\end{figure}

\begin{figure}
\includegraphics[width=0.45\textwidth,height=0.3375\textwidth]{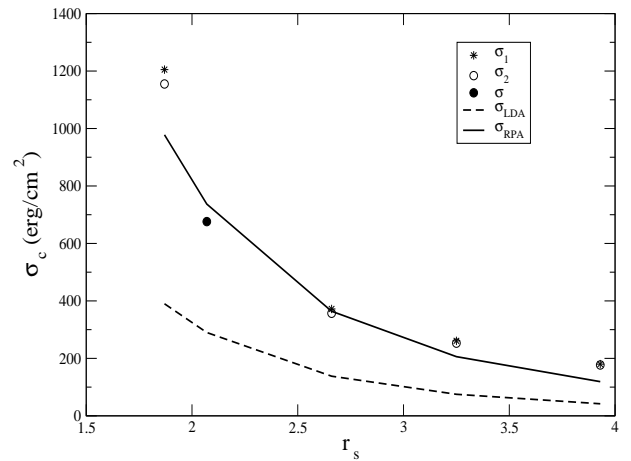}
\caption{Correlation contribution to the surface energy. Stars and open circles
represent DMC correlation surface energies obtained by combining the
Acioli-Ceperley fixed-node slab energies with the fixed-node bulk energies
of Eqs.~(\ref{par}) and (\ref{par2}), respectively. The solid circle represents
the DMC correlation surface energy obtained at $r_s=2.07$ by combining the
Acioli-Ceperley fixed-node slab energy with the corresponding fixed-node bulk
energy $\varepsilon^{bulk}=-0.01482\,{\rm Ry}$ of Ceperley and
Alder.\cite{ca2} Dashed and solid lines represent LDA and RPA calculations,
respectively.}\label{fig2}
\end{figure}

Fig.~1 exhibits the surface energies of Table~I, as a function of $r_s$. At
the highest densities the agreement between DMC and RPA surface energies is
good. At the smallest densities ($r_s\ge 3.25$) the corrected DMC
surface energies are still larger than their RPA counterparts. Nevertheless, we
note that DMC surface energies are very sensitive to small uncertainties in
the parametrization of the bulk energy. Therefore, if one is to quantitatively
account for jellium surface energies, both DMC slab and bulk energies entering
Eq.~(\ref{sigma}) should be computed on the same footing.

Small differences between the DMC and RPA calculations come from the
correlation contribution to the surface energy. We have carried out
calculations of exact kinetic ($\sigma_s$), electrostatic ($\sigma_{es}$), and
exchange ($\sigma_x$) surface energies for jellium slabs with
$L=7.21\,(r_s\,a_0)$,
and we have defined the correlation contribution to the DMC surface energy as
\begin{equation}
\sigma_c=\sigma-(\sigma_s+\sigma_{es}+\sigma_x).
\end{equation}
The results we have obtained from our analysis of the Acioli-Ceperley DMC
slab energies are shown in Fig.~2 (see also Table~I), together with the LDA and
RPA correlation energies that we have obtained for the same jellium slabs.
This figure clearly shows that combining fixed-node slab energies with their
fixed-node bulk counterparts yields DMC correlation surface energies in
reasonable agreement with the results obtained in the RPA. While the LDA
underestimates considerably the correlation surface energy for all electron
densities, it overestimates the exact exchange surface energy.
Large and opposite separate corrections to the LDA for exchange and
correlation largely compensate, as discussed before,\cite{pitarke1,perdew} and
yield LDA surface energies that are close to their non-local counterparts.

In conclusion, assuming that the fixed-node error in the slab and the
bulk calculations cancel out, the DMC data reported in
Ref.~\onlinecite{acioli} yields surface energies that are considerably closer
to the LDA and in reasonable agreement with those obtained in the RPA.
Nevertheless, at the smallest densities the corrected DMC surface energies are
still larger than their RPA counterparts; at these densities, they are also
larger than the jellium surface energies extracted from DMC calculations for
jellium spheres,\cite{ballone} which are found to be close to the LDA.\cite{z}
Surface energies are extremely sensitive to little uncertainties in both slab
and bulk energies; hence, in order to quantitatively account for the impact of
nonlocal xc effects on the surface energy one needs to pursue DMC evaluations
of both slab and bulk energies on the same footing.

Partial support by the University of the Basque Country, the
Basque Unibertsitate eta Ikerketa Saila, and the Spanish MCyT is acknowledged.
I thank J. P. Perdew for stimulating discussions and D. M. Ceperley for
correspondence related to the unpublished uniform-gas fixed-node data for
$r_s=2.07$.


\begin{thebibliography}{100}
\bibitem{note0} The difference between this surface energy and the surface
energy of $-558\,{\rm erg/cm^2}$ that we have obtained from the
Acioli-Ceperley slab energy (see caption of Table~I) is partially due to the
difference
in the thickness of the jellium  slabs considered in Refs.~\onlinecite{acioli}
and ~\onlinecite{li}.
\bibitem{note1} Due to oscillatory quantum-size effects,
differences between the finite-slab surface energies and those reported before
for the semi-infinite jellium\cite{zhang,pitarke1} are larger for
$L=8.52\,(r_s\,a_0)$
than in the case of $L=7.21\,(r_s\,a_0)$.
\end{thebibliography}
\end{document}